\documentclass[10pt,twocolumn,letterpaper]{article}

 \usepackage[pagenumbers]{cvpr} % To force page numbers, e.g. for an arXiv version

\usepackage{makecell}

\definecolor{cvprblue}{rgb}{0.21,0.49,0.74}
\usepackage[pagebackref,breaklinks,colorlinks,allcolors=cvprblue]{hyperref}

\def\paperID{} % *** Enter the Paper ID here
\def\confName{CVPR}
\def\confYear{2026}

\newcommand{\mname}{{BATISNet}}

\title{BATISNet: Instance Segmentation of Tooth Point Clouds\\ with Boundary Awareness}

\author{Yating Cai$^{1}$, \, Yanghui Xu$^{1}$, \, Zehua Hu$^{1}$, \, Jiazhou Chen$^{1,2,*}$, \, Jing Huang$^{3}$ \\
$^{1}$Zhejiang University of Technology \\ 
$^{2}$Zhejiang Key Laboratory of Visual Information Intelligent Processing \\ 
$^{3}$Zhejiang Gongshang University \\
$^*$ Corresponding author: cjz@zjut.edu.cn
}

\begin{document}
\maketitle
\begin{abstract}
Accurate segmentation of the tooth point cloud is of great significance for diagnosis clinical assisting and treatment planning. Existing methods mostly employ semantic segmentation, focusing on the semantic feature between different types of teeth. However, due to the tightly packed structure of teeth, unclear boundaries, and the diversity of complex cases such as missing teeth, malposed teeth,  semantic segmentation often struggles to achieve satisfactory results when dealing with complex dental cases. To address these issues, this paper propose BATISNet, a boundary-aware instance network for tooth point cloud segmentation. This network model consists of a feature extraction backbone and an instance segmentation module. It not only focuses on extracting the semantic features of different types of teeth but also learns the instance features of individual teeth. It helps achieve more robust and accurate tooth instance segmentation in complex clinical scenarios such as missing teeth and malposed teeth. Additionally, to further enhance the completeness and accuracy of tooth boundary segmentation, a boundary-aware loss function is designed to specifically supervise the boundary segmentation between instances. It mitigates effectively tooth adhesion and boundary ambiguity issues. Extensive experimental results show that BATISNet outperforms existing methods in tooth integrity segmentation, providing more reliable and detailed data support for practical clinical applications.
\end{abstract}    
\section{Introduction}
\label{sec:intro}

With the rapid advancement of digital technology, computer-based approaches have been increasingly integrated into medical practice. In digital dentistry, intraoral scanning provides the foundation for many modern treatments, making the segmentation of these models a key challenge. Computer-assisted separation of teeth and gingiva offers critical data for clinical decision-making and treatment planning in orthodontics, prosthodontics, implantology, and related fields.

However, accurate tooth instance segmentation in intraoral scan models remains a significant challenge. Historically, the segmentation and classification of intraoral scans have relied heavily on manual annotation or traditional manually designed geometric feature-based methods~\cite{sun2020automatic, sun2020tooth, zhang2020automatic}. Although the introduction of deep neural networks has considerably advanced this field, several limitations persist. First, learning-based approaches typically require extensive manually annotated datasets, which are time-consuming and labor-intensive to produce. Second, owing to the small gaps between teeth and insufficient scanning accuracy, many methods exhibit reduced accuracy in delineating precise tooth boundaries. Third, the high variability in real-world dental conditions poses additional difficulties; most existing techniques are primarily designed for well-aligned dentition and thus perform poorly in clinically complex cases.

Most existing approaches formulate tooth segmentation as a semantic segmentation task, where each tooth is assigned a distinct category label—such as incisors, canines, premolars, and molars. These methods typically extract high-level features, including shape and semantic information, from input data such as point coordinates, normal vectors~\cite{zhang2021tsgcnet, lian2020deep} or curvatures~\cite{sun2020automatic, xiong2023tsegformer} of dental point clouds. These features are then used to predict the category of each point. However, due to the densely packed arrangement of teeth, limited scanning resolution along tooth boundaries, and morphological variations caused by dental diseases, such semantic segmentation methods often struggle to delineate each tooth instance accurately and completely. This is particularly evident in challenging clinical cases involving missing teeth, malpositioned teeth, or tightly packed dental structures.

To address these challenges, we propose \mname\, a tooth instance segmentation network that formulates tooth segmentation as an instance segmentation task. The model takes point coordinates and normal vectors as input and employs a PointMLP~\cite{ma2022rethinking} backbone to extract discriminative features from raw tooth point clouds. These features are subsequently processed by an instance segmentation module to generate instance-wise embeddings, enabling the precise separation of individual tooth instances. Additionally, a boundary-aware loss function is introduced to supervise the training process, enhancing both the structural completeness of segmented teeth and the geometric accuracy of their boundaries.

The main contributions are summarized as follows:
\begin{itemize}
    \item An instance segmentation network named \mname\ is proposed for tooth point clouds. By integrating the efficient feature extraction capability of PointMLP with the instance-wise feature embedding of an instance segmentation module, the network effectively handles clinically complex cases, including those with missing teeth, wisdom teeth, malposed teeth, and other anatomical variations.
    \item A boundary-aware loss function is designed to enhance the discriminative ability of instance boundaries, achieving high segmentation accuracy with an end-to-end optimization.
    \item Plenty of experiments on open datasets demonstrate that \mname\ outperforms SOTA methods, and ablation studies show the validity of major modules of the proposed network.
\end{itemize}
\section{Related Works}
\label{sec:related}

%-------------------------------------------------------------------------
\subsection{3D tooth semantic segmentation}
Traditional approaches for segmenting tooth intraoral scan models, including those based on curvature~\cite{sun2020automatic, xiong2023tsegformer}, contour lines~\cite{sinthanayothin2008orthodontics, yaqi2010computer}, or predefined geometric criteria~\cite{zou2015interactive}, predominantly rely on handcrafted feature design. These methods often suffer from limited efficiency and poor generalization capability. With the rapid advancement of deep learning, an increasing number of studies have explored data-driven approaches for 3D tooth segmentation.~\cite{jana2023critical,wang2024weakly,10654940,10230650}. Most of these learning-based methods formulate the task as a semantic segmentation problem, typically adopting a predefined label set of 16 tooth categories and one gingival class, and performing per-point classification on the tooth point cloud.

Lian et al. proposed a multi-scale MeshSNet for end-to-end segmentation of tooth mesh data~\cite{lian2019meshsnet}. Hao et al. proposed a fully automatic segmentation and detection system capable of generating fine-grained segmentation and assessing whether the results require correction~\cite{hao2022toward}. Zhang et al. proposed TSGCNet, which innovatively adopted a dual-stream network, taking the coordinates and normal vectors of tooth point clouds as input to extract high-level semantic features and combine multi-view complementary information for tooth segmentation~\cite{zhang2021tsgcnet}. Zheng et al. proposed TeethGNN, a GNN-based framework that autonomously learns discriminative geometric representations from raw dental data, significantly outperforming traditional manual feature engineering approaches~\cite{zheng2022teethgnn}. Xiong et al. proposed TSegFormer, which achieves clinically viable segmentation accuracy through a Transformer architecture and a curvature-based geometric guidance loss~\cite{xiong2023tsegformer}. 

However, this semantic segmentation paradigm, fundamentally constrained by its predefined and fixed set of tooth classes, relies on ideal dental arrangements and completeness. This limitation severely compromises its performance in segmenting atypical cases, including missing or malposed teeth.

%-------------------------------------------------------------------------
\subsection{3D tooth instance segmentation}
Another research direction formulates the tooth segmentation task as instance segmentation, aiming to segment individual tooth instances from the input intraoral scan model while classifying each tooth instance~\cite{Xi20253DDM}. Zanjani et al. proposed an instance segmentation method that first predicts voxel bounding boxes and then segments the tooth foreground within these boxes~\cite{zanjani2021mask}. Cui et al. introduced a two-stage TSegNet network, which initially predicts tooth centroids with distance awareness and subsequently incorporates a confidence-aware attention module to achieve tooth instance segmentation~\cite{cui2021tsegnet}. Tian Y et al. first predict candidate center points for each tooth, then aggregate points belonging to the same tooth through affinity learning, and finally achieve precise segmentation of individual teeth via clustering~\cite{tian20223d}. Qiu et al. enhanced 3D tooth instance segmentation under weak annotations by incorporating dental arch shape as a geometric prior~\cite{qiu2022darch}.

\begin{figure*}[!ht]
\centering
    \includegraphics[width=0.95\linewidth]{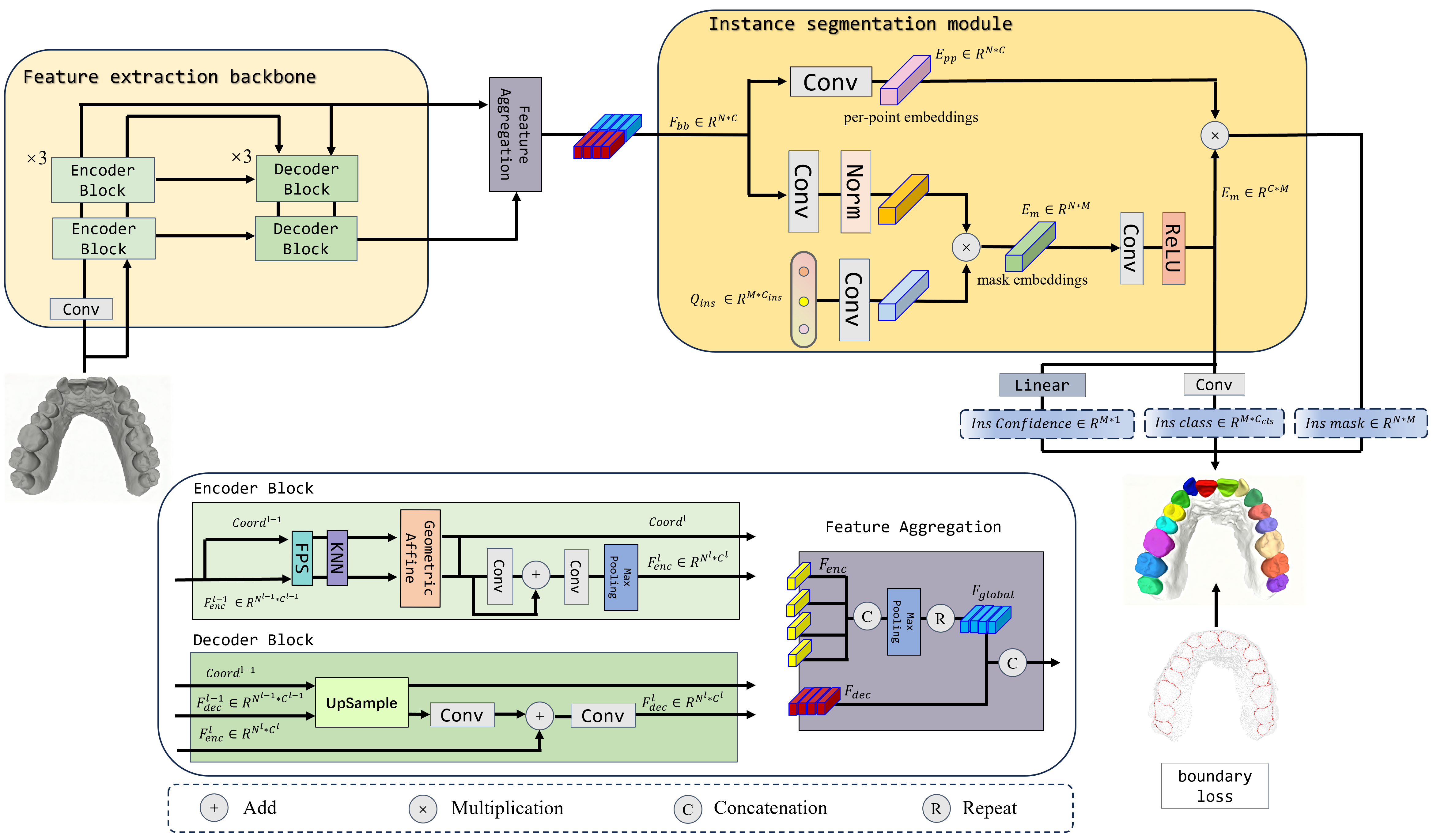}
    \caption{Overview of \mname. The architecture comprises three main components: (1) a feature extraction backbone with a U-Net-shaped encoder-decoder for hierarchical point cloud feature extraction; (2) an instance mask-aware segmentation module that leverages learnable dental instance queries and the extracted features to perform simultaneous tooth instance segmentation and classification; (3) a boundary-aware loss that enhances segmentation accuracy at inter-tooth boundaries.
        \label{fig:overview}
        \vspace{0.5em}
    }
\end{figure*}

In contrast, instance segmentation methods offer a more flexible alternative by naturally accommodating a variable number of tooth instances and leveraging strong instance-aware representations. Nevertheless, most existing instance-based approaches rely heavily on accurate preliminary predictions, such as tooth bounding boxes or centroids, to guide the segmentation process. The foreground-background separation within these localized regions is often performed without sufficient global contextual awareness, limiting their robustness. The method proposed in this work addresses these issues through an instance-aware mask embedding mechanism that effectively integrates both local fine-grained features and global semantic information. Importantly, our approach operates in a proposal-free manner, eliminating the dependency on the quality of intermediate candidate sets such as bounding boxes or centroids.

%-------------------------------------------------------------------------
\subsection{3D tooth boundary detection}
Accurate boundary detection plays a vital role in distinguishing adjacent instances and mitigating issues such as adhesion or mis-segmentation between neighboring structures, thereby substantially improving the overall performance of instance segmentation. This capability is especially critical in applications involving densely packed and morphologically complex anatomical structures, such as teeth and biological cells, where precise delineation of instance boundaries directly influences segmentation accuracy and clinical applicability. 

In the field of point cloud boundary detection, Hu et al. proposed a novel two-stream fully convolution network to address 3D semantic edge detection~\cite{hu2020jsenet}. By incorporating a joint refinement module and a specific loss function, their network produces semantic segmentation results with improved boundaries. Gong et al. introduced a Boundary Prediction Module (BPM) to predict boundary points and a boundary-aware Geometric Encoding Module (GEM) to encode geometric information while performing discriminative aggregation of features within local neighborhoods~\cite{marin2019efficient}. However, both studies involve complex modules or local aggregation for explicit boundary prediction, which increases model complexity. 

In the field of dental boundary detection, Xu et al. using graph optimization to enhance boundary segmentation between teeth and between teeth and gums~\cite{xu20183d}. Yuan et al. introduced a tooth boundary detection method based on principal curvature analysis~\cite{yuan2020tooth}. These two boundary detection methods rely on manually determined parameters and features, and cannot be trained in an end-to-end manner, resulting in both low efficiency and poor generalization. Xiong et al. proposed a curvature-based geometrically guided boundary loss to optimize tooth boundary detection~\cite{xiong2023tsegformer}. However, such curvature-based methods tend to over-prioritize high-curvature regions at the expense of accurately identifying boundaries in low-curvature areas. In contrast, our proposed boundary loss provides a more comprehensive solution that enhances detection accuracy across all boundary types, while maintaining a lightweight design and enabling end-to-end optimization.

\section{Methodology} 
\label{sec:method}

In this paper, a boundary-aware instance segmentation network is proposed for 3D tooth point cloud segmentation. As illustrated in Figure~\ref{fig:overview}, the framework consists of three key components: a modified U-Net backbone based on PointMLP~\cite{ma2022rethinking} (Sec. 3.2), which extracts both local and global features from complex point clouds; a mask-aware instance segmentation module (Sec. 3.3), which simultaneously performs instance segmentation, semantic classification, and confidence prediction for each tooth; and a boundary loss function (Sec. 3.4), which enhances the accuracy of instance boundary detection. The following sections provide a detailed description of these components.

%-------------------------------------------------------------------------
\subsection{Feature extraction backbone}
To accommodate the characteristics of irregular point distribution, densely packed instances, and complex boundaries in dental point clouds, a PointMLP-based feature extraction backbone is constructed to effectively extract dental features. We transform the original PointMLP network into a U-Net architecture, comprising an encoder and a decoder, and utilize skip connections to integrate features from corresponding layers of the encoder and decoder. Given an input point cloud $X\in \mathbb{R}^{N \times C_{in}}$, where N denotes the number of points, the input features include point coordinates, normal vectors, and the coordinates and normal vectors of the three points in the point cloud's corresponding grid. The input features are mapped to a high-dimensional space through a 1D convolutional layer and a normalization layer.

Each encoder block begins with Farthest Point Sampling (FPS) operation on the input point cloud $X$ for downsampling, then computes K-nearest neighbors for the subsampled points. Record $l$ as the layer number of the encoder and decoder, for the neighborhood features $F_{enc}^{l-1} \in \mathbb{R}^{N^{l-1} \times K \times C^{l-1}}$ of each point, a geometric affine transformation module enhances the adaptive representation capability of geometric features across different local regions:
\begin{equation}
    \hat{F} = \alpha \cdot \frac{F - \mu}{\sigma + \epsilon} + \beta
\end{equation}
Here, $\mu$ and $\sigma$ represent the mean and standard deviation of the features of the K-nearest neighboring points, respectively, $\alpha$ and $\beta$ are learnable parameters, and $\epsilon$ is a small constant (e.g., 1e-5). The transformed local features are then processed through 1D convolution and residual connections for feature extraction. Finally, the features $F_{enc}^{l} \in \mathbb{R}^{N^l \times K \times C^l}$ are aggregated along the K dimension using max pooling, producing the output features $F_{enc}^{l} \in \mathbb{R}^{N^l \times C^l}$, along with the coordinates $Coord^l$ of the sampled points after FPS downsampling.

In each decoder block, the output features from the previous decoder layer are upsampled and interpolated to obtain higher-resolution point cloud features, which are then fused with the output features from the corresponding encoder layer. The output feature of the decoder layer $l$ is $F_{dec}^{l} \in \mathbb{R}^{N^l \times C^l}$.

To better integrate the global features of intraoral scan models with the local regional features between each point and its K-nearest neighbors, thereby enhancing the feature extractor's capacity to represent both global and local information of the intraoral scans, the proposed method employs a global-local feature aggregation module. This module first concatenates the output features $F_{enc}^l \in \mathbb{R}^{N^l \times C^l}$ from multiple encoder layers along the feature channel dimension then performs max pooling along the point dimension $N$ to obtain global features $F_global^l \in \mathbb{R}^{1 \times C^l}$. These global features are replicated $N$ times and subsequently concatenated with the output features $F_{dec}$ from the last decoder layer along the feature channel dimension, yielding the final output feature $F_{bb} \in \mathbb{R}^{N \times C}$ of the feature extraction backbone. The formulation is as follows:
\begin{equation}
    F_{bb} =Cat\Big(F_{dec}, Repeat\big((MaxPooling(Cat(F_{enc}^l))\big)\Big)    
\end{equation}

%-------------------------------------------------------------------------
\subsection{Instance segmentation module}
Following the feature extraction backbone, an instance-aware segmentation module is incorporated to achieve precise identification and segmentation of tooth instances in the point cloud. Inspired by 2D image instance segmentation approaches such as OneFormer~\cite{jain2023oneformer} and Mask2Former~\cite{cheng2022masked}, we formulate the tooth segmentation task as a binary mask prediction and mask classification problem. The paper designs a simplified Transformer decoder~\cite{vaswani2017attention} module to efficiently extract per-point embeddings and generate mask embeddings with instance-aware capabilities.

Similar to the Transformer decoder, with a maximum number of tooth instances denoted as M, the method first maps a learnable tooth instance query $Q_{ins} \in \mathbb{R}^{M \times C_{ins}}$ and the point cloud features $F_{bb} \in \mathbb{R}^{N \times C}$ extracted by the feature extraction backbone through separate convolutional layers before multiplying them, yielding the instance-aware mask embedding $E_m \in \mathbb{R}^{M \times N}$. The method then employs three parallel branches for instance mask prediction, mask category prediction, and mask confidence prediction. In the tooth instance mask prediction branch, the per-point embedding $E_{pp} \in \mathbb{R}^{N \times C}$ – obtained by applying 1D convolution to point cloud features $F_{bb} \in \mathbb{R}^{N \times C}$ – is multiplied with the mask embedding $E_m$ to generate binary maps of candidate instance masks. The mask category prediction branch utilizes a 1D convolutional as linear classifier to predict the category of each mask based on $E_m$. For mask confidence prediction, a simple linear layer followed by a activation layer estimates the confidence score for each mask. Finally, the outputs from these three branches are combined, and non-maximum suppression is applied to select the most reliable tooth instance masks.

%-------------------------------------------------------------------------
\subsection{Boundary loss function}

To further improve the segmentation accuracy at instance boundaries, this paper proposes an Instance Boundary Loss ($L_{ibl}$) specifically designed for optimizing boundary segmentation between instances. The most challenging aspect of instance segmentation tasks often lies in the adjoining regions between different instances, where conventional loss functions demonstrate limited discriminative capability. Therefore, $L_{ibl}$ is computed exclusively on the set of boundary points $B$ to emphasize the model's focus on boundary regions. Specifically, we define it as:

\begin{equation}
    L_{ibl} = -\frac{1}{|B|} \sum_{i \in B} \sum_{c=1}^{C} \big( (1 - p_{i,c})^\gamma \cdot y_{i,c} \cdot \log(p_{i,c}) \big)  
\end{equation}

\noindent where $B$ represents the set of all boundary points, $C$ denotes the number of tooth categories, $p_{i,c}$ indicates the probability of point $i$ being predicted as class $c$ in the instance mask $m$, $y_{i,c}$ represents the ground truth label of point $i$, and $\gamma$ serves as the modulating factor of the Focal Loss, empirically set to 2. By assigning higher weights to hard-to-classify boundary points, this loss guides the model to focus on the boundaries between instances, effectively mitigating issues such as instance adhesion and boundary ambiguity.
The overall loss function of this method is:

\begin{equation}
    L=\lambda_{cls} \cdot L_{cls}+\lambda_{mask} \cdot L_{mask}+\lambda_{obj} \cdot L_{obj}+\lambda_{ibl} \cdot L_{ibl}  
\end{equation}

\noindent where $L_{cls}$ represents the Cross Entropy Loss that supervises the tooth classification task; $L_{obj}$ denotes the object confidence loss, which selects high-confidence tooth instance segmentation candidates; $L_{mask}$ constitutes the mask loss, comprising both binary cross-entropy loss and Dice Loss~\cite{sudre2017generalised} that directly optimizes the IoU between predicted masks and ground truth masks. The hyperparameter $\lambda_{ibl}$ is set to 0.006. 

Since the number of predicted instances may differ from the number of ground truth instances, instance mask matching becomes necessary during loss computation. This paper employs the Hungarian Algorithm~\cite{kuhn2005hungarian} to establish optimal matching between predicted instances and ground truth instances, thereby ensuring each predicted instance corresponds to the most suitable ground truth instance, effectively enhancing segmentation and classification accuracy.
Through the above design, the model can jointly optimize instance mask segmentation, instance category prediction, and instance object confidence prediction, achieving high-precision instance recognition and segmentation of tooth point clouds.

%-------------------------------------------------------------------------
\subsection{Post processing}
Graph Cut has been successfully applied to automated optimization of image segmentation results~\cite{chen2017deeplab, cai2017pancreas}. To further improve tooth boundary segmentation accuracy and avoid isolated erroneous predictions, this paper introduces Graph Cut post-processing~\cite{lian2019meshsnet, liu2022hierarchical} based on the network's predictions, aiming to produce more coherent segmentation results with smoother boundaries. Treating each point as a graph node, the method constructs graph edges using K-nearest neighborhoods. The Graph Cut method obtains optimized point cloud labels by minimizing the global energy $E(L)$, with the overall energy function defined as:

\begin{equation}
    E(L) = \sum_{p \in P} U_p(L_p) + \lambda \cdot \sum_{(p,q) \in E} w_{pq} \cdot \mathbb{I}[L_p \neq L_q]
\end{equation}

\noindent where the first term is the data term, and the second term is the smoothness term, $\lambda$ is the smoothing weight, set to 2. 
In the data term, $U_p(c)$ represents the cost of assigning point $p$ to category $c$, defined as:

\begin{equation}
    U_p(c) = -\log(P(p, c) + \epsilon)
\end{equation}

\noindent where $P \in \mathbb{R}^{N \times(C+1))}$ represents the probability distribution of each point output by \mname, and $\epsilon$ is a small constant (e.g., $e^{-5}$). 

The second term in $E(L)$ is the smoothness term, where $w_{pq}$ is defined as:

\begin{equation}
    w_{pq} = \frac{(1 - d_{pq}) \cdot (1 + nsim_{pq})}{2}
\end{equation}

\noindent where $d_{pq}$ represents the normalized distance between the two nodes of the edge; and $nsim_{pq}$ denotes the cosine similarity of the normal vectors at points $p$ and $q$.

\begin{figure}[!ht]
\centering
    \includegraphics[width=0.9\linewidth]{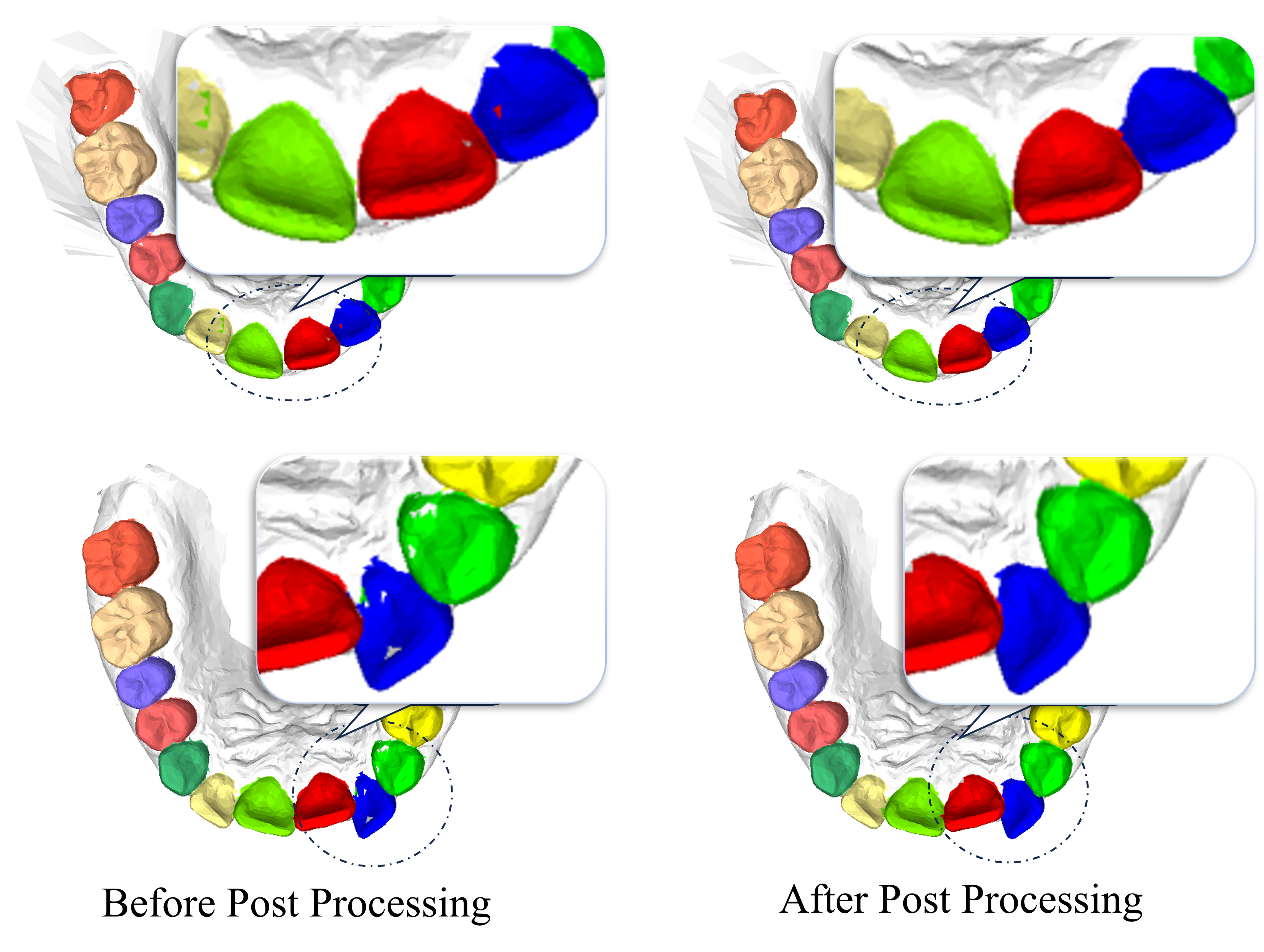}
    \caption{Segmentation results before and after graph cut optimization.
        \label{fig:graphcut}
        \vspace{0.5em}
    }
\end{figure}

Figure~\ref{fig:graphcut} also presents a comparison of the visualization results before and after graph cut processing. It can be observed that some minor defects disappear after graph cut processing, which validates the effectiveness of graph cut.

\section{Experiments} 
\label{sec:experiments}

\begin{figure*}[!ht]
\centering
    \includegraphics[width=\linewidth]{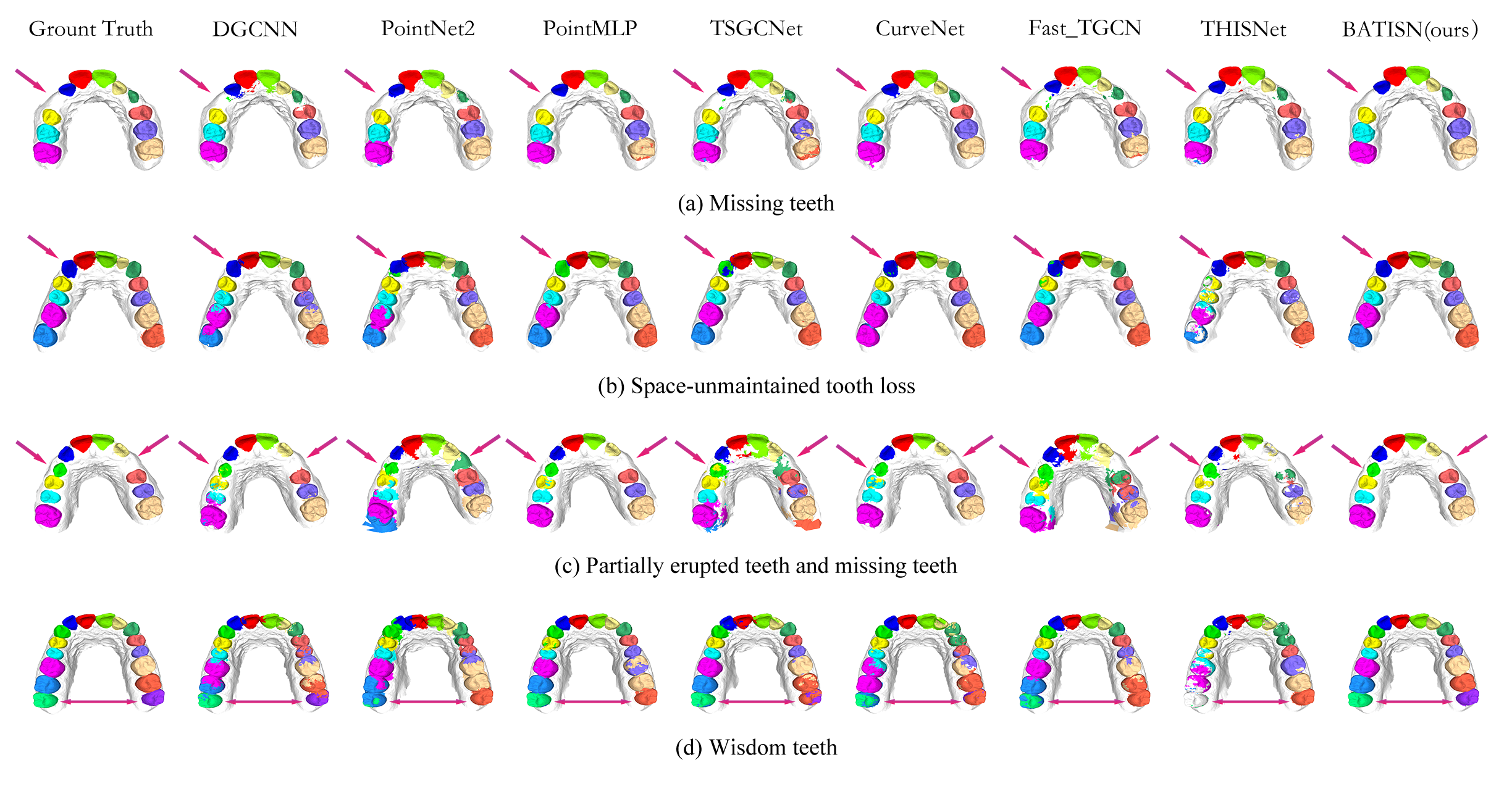}
    \caption{Visualization of representative segmentation results generated by eight competing methods and our method. Specifically, the first row presents representative results of missing  teeth; the second row shows representative results of the scenario with space-unmaintained tooth loss, where there is no positional gap but one canine tooth (T3) is essentially missing; the third row illustrates representative results of partially erupted teeth and missing teeth; the fourth row displays representative results of scenarios with wisdom teeth.
        \label{fig:compare}
        \vspace{0.5em}
    }
\end{figure*}

\subsection{Experimental settings}
\begin{figure}[!ht]
\centering
    \includegraphics[width=0.9\linewidth]{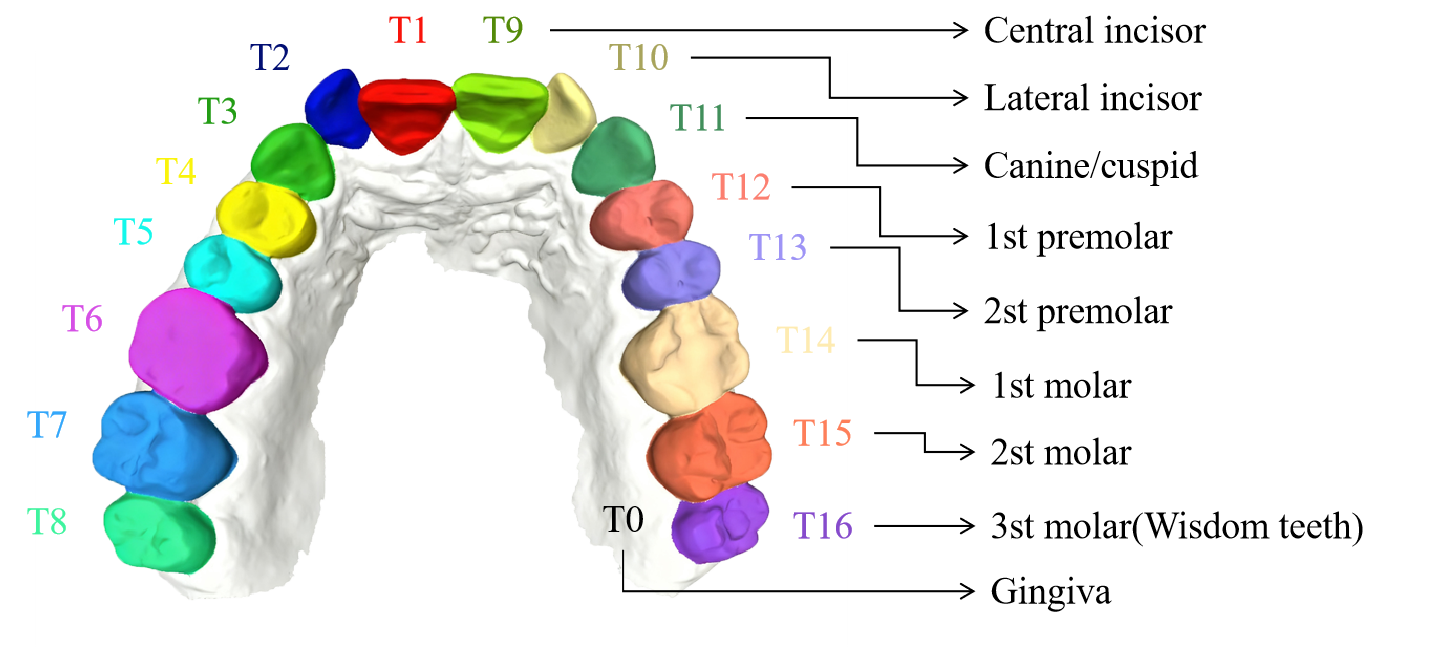}
    \caption{The index of teeth. All teeth, including wisdom teeth, are labeled into 16 categories. The teeth with relative positions on both sides belong to the same category.
        \label{fig:index_of_teeth}
        \vspace{0.5em}
    }
\end{figure}

\noindent \textbf{The datasets.} The publicly available dataset from the MICCAI Challenge~\cite{ben20233dteethseg} is utilized in our experiments. A total of 254 maxillary models are selected for training and 125 for testing. All models are uniformly downsampled to 16,000 points while preserving their original topological structures. The dataset includes annotations for 16 tooth categories (including wisdom teeth), with gingival tissue labeled as background. For each intraoral scan, individual tooth instances are annotated from T1 to T16, and gingiva is labeled as T0, as illustrated in Figure~\ref{fig:index_of_teeth}. The proportion of models containing wisdom teeth is relatively low in both training and test sets, each accounting for approximately 5\% of the respective dataset.

\noindent \textbf{Implementations.} Both training and testing phases are implemented on an NVIDIA RTX 3090 GPU using PyTorch. We set the batch size to 6, the training epochs to 300, and the initial learning rate to $e^{-3}$.

\noindent \textbf{Evaluation metrics.} To evaluate the segmentation performance of our method, we employ several widely used metrics: Accuracy (ACC), Intersection over Union (IoU), and the Dice coefficient. Among these, Accuracy reflects the overall point‑wise classification correctness, while IoU and Dice measure the spatial overlap and similarity between predicted results and ground‑truth annotations, respectively. Furthermore, we incorporate the Average Precision (AP) metric to specifically assess instance‑level segmentation quality, enabling a more holistic evaluation of both detection completeness and boundary accuracy.

\subsection{Comparison with SOTA methods} 

\begin{table}[h]
\centering
\caption{Comparison of segmentation results between our method and seven competing methods on OA, mAcc, and mIoU. In the table, Cat. denotes the category to which the method belongs, while Sem. and Ins. represent Semantic Segmentation and Instance Segmentation, respectively.}
\label{tab:comp_accuracy}
\begin{tabular}{c|c|ccc}
\textbf{Methods} & \textbf{Category} & \textbf{OA} & \textbf{mACC} & \textbf{mIOU} \\
\hline
%DGCNN\cite{wang2019dynamic} & Sem. & 93.60\% & 85.80\% & 78.88\% \\
%PointNet2\cite{qi2017pointnet++} & Sem. & 88.28\% & 80.10\% & 68.03\% \\
%PointMLP\cite{ma2022rethinking} & Sem. & 94.78\% & 86.84\% & 80.29\% \\
%TSGCNet\cite{zhang2021tsgcnet} & Sem. & 94.98\% & 88.90\% & 82.47\% \\
%CurveNet\cite{muzahid2020curvenet} & Sem. & 94.16\% & 87.04\% & 79.20\% \\
%Fast-TGCN\cite{li2024fine} & Sem. & 95.03\% & 88.33\% & 82.11\% \\
%THISNet\cite{li2023thisnet} & Ins. & 92.93\% & 82.34\% & 76.28\% \\
%Ours & Ins. & \textbf{95.20\%} & \textbf{90.19\%} & \textbf{84.42\%}
DGCNN & Semantic & 93.60\% & 85.80\% & 78.88\% \\
PointNet2 & Semantic & 88.28\% & 80.10\% & 68.03\% \\
PointMLP & Semantic & 94.78\% & 86.84\% & 80.29\% \\
TSGCNet & Semantic & 94.98\% & 88.90\% & 82.47\% \\
CurveNet & Semantic & 94.16\% & 87.04\% & 79.20\% \\
Fast-TGCN & Semantic & 95.03\% & 88.33\% & 82.11\% \\
THISNet & Instance & 92.93\% & 82.34\% & 76.28\% \\
Ours & Instance & \textbf{95.20\%} & \textbf{90.19\%} & \textbf{84.42\%}
\end{tabular}
\end{table}

In this section, we compare our method with other SOTA approaches including THISNet~\cite{li2023thisnet}, TSGCNet~\cite{zhang2021tsgcnet}, PointMLP~\cite{ma2022rethinking}, PointNet2~\cite{qi2017pointnet++}, CurveNet~\cite{muzahid2020curvenet}, Fast-TGCN~\cite{li2024fine}, and DGCNN~\cite{wang2019dynamic}. To ensure a fair comparison, all models are trained and tested under the same conditions using their official implementations.

The experimental results are shown in Table~\ref{tab:comp_accuracy}. Our proposed network consistently achieves state-of-the-art performance across all evaluation metrics, including OA, mACC, and mIoU. Specifically, compared to the baseline PointMLP backbone, \mname\ yields a notable improvement of 3.35\% in mACC and 4.13\% in mIoU. Furthermore, when evaluated against THISNet, another instance-based segmentation method, our method exhibits clear performance gains in all metrics.

Figure~\ref{fig:compare} presents a visual comparison between the segmentation results of our method and other approaches, including cases with missing teeth and wisdom teeth. The first row shows the segmentation results for cases with missing teeth, where our method accurately identifies the vacant positions and achieves correct segmentation on adjacent teeth. The second row demonstrates the segmentation results for space-unmaintained tooth loss, a scenario where there is no positional gap but one canine tooth (T3) is essentially missing. Our method successfully assigns correct labels to tooth instances even at the missing position. The third row presents the segmentation results for missing teeth and partially erupted teeth cases, where only our method achieves precise segmentation despite the combined challenges of missing and partially erupted teeth. The fourth row displays the wisdom tooth segmentation results, indicating our method's strong generalization capability in segmenting less frequent tooth types. Across these complex and low frequency clinical scenarios, our method consistently delivers superior segmentation performance. This advantage originates from our instance mask based prediction approach, which fundamentally differs from traditional point wise semantic classification. Our framework remains robust to variations in tooth instance count while demonstrating accurate perception of complete tooth instance structures. In contrast, semantic segmentation based methods frequently produce inconsistent semantic labels within individual tooth instances, particularly in these challenging regions.

In addition to the semantic segmentation metrics, we employ Average Precision (AP), a standard evaluation criterion in instance segmentation tasks. While semantic metrics reflect the overall segmentation quality of intraoral scans, clinical applications require precise delineation of individual teeth. AP@T specifically quantifies segmentation accuracy at the tooth instance level, where the parameter T denotes the IoU threshold between predicted and ground truth instances. A prediction is considered correct when its IoU exceeds threshold T. The mean Average Precision (mAP) is computed as the average AP across 10 IoU thresholds ranging from 0.5 to 0.95 with a step size of 0.05. For fair comparison with semantic segmentation based methods, each semantic label is treated as a separate instance during AP calculation, with gingival labels excluded as background. The complete results are presented in Table~\ref{tab:comp_ap}. Our method achieves the highest scores in both mAP and AP across all IoU thresholds. Compared to THISNet, another instance segmentation approach, our method maintains at least a 3\% performance improvement, with the advantage becoming more substantial at higher IoU thresholds. Significant improvements are also observed when compared to our backbone network, PointMLP.

\begin{table}[h]
\centering
\caption{Comparison of existing SOTA methods at various Average Precision thresholds.}
\label{tab:comp_ap}
%\begin{tabular}{c|ccccc|c}
%\textbf{Method} &  \textbf{AP$_{50}$} & \textbf{AP$_{60}$} & \textbf{AP$_{70}$} & \textbf{AP$_{80}$} & \textbf{AP$_{90}$} & \textbf{mAP} \\
%\hline
%DGCNN &  86.40\% & 79.13\% & 76.90\% & 71.94\% & 59.12\% & 74.70\% \\
%PointNet2 &  81.55\% & 77.73\% & 63.03\% & 29.80\% & 0.18\% & 50.46\% \\
%PointMLP &  80.46\% & 80.58\% & 76.86\% & 74.31\% & 63.89\% & 75.22\% \\
%TSGCNet & 86.14\% & 81.33\% & 80.49\% & 75.92\% & 63.95\% & 77.57\% \\
%CurveNet &  81.25\% & 77.94\% & 75.29\% & 70.57\% & 61.07\% & 73.22\% \\
%Fast-TGCN &  82.68\% & 80.67\% & 80.88\% & 74.85\% & 62.03\% & 76.22\% \\
%THISNet &  85.46\% & 80.34\% & 78.87\% & 73.64\% & 56.77\% & 75.02\% \\
%Ours &  \textbf{88.89\%} & \textbf{87.62\%} & \textbf{85.00\%} & \textbf{80.57\%} & \textbf{67.57\%} & \textbf{81.93\%} 
\begin{tabular}{c|ccc|c}
\textbf{Methods} &  \textbf{AP@0.5} & \textbf{AP@0.7} &  \textbf{AP@0.9} & \textbf{mAP} \\
\hline
DGCNN     &  86.40\% & 76.90\% & 59.12\% & 74.70\% \\
PointNet2 &  81.55\% & 63.03\% & 0.18\% & 50.46\% \\
PointMLP  &  80.46\% & 76.86\% & 63.89\% & 75.22\% \\
TSGCNet   & 86.14\%  & 80.49\% & 63.95\% & 77.57\% \\
CurveNet  &  81.25\% & 75.29\% & 61.07\% & 73.22\% \\
Fast-TGCN &  82.68\% & 80.88\% & 62.03\% & 76.22\% \\
THISNet   &  85.46\% & 78.87\% & 56.77\% & 75.02\% \\
Ours &  \textbf{88.89\%} & \textbf{85.00\%} & \textbf{67.57\%} & \textbf{81.93\%} 
\end{tabular}
\end{table}

\begin{table}[h]
\centering
\caption{Comparison of existing SOTA methods in complex tooth dataset. This data subset containing challenging cases (e.g., missing teeth and malposed teeth.) is selected from the original validation set.} 
\label{tab:comp_special}
\begin{tabular}{c|cccc}
\textbf{Methods}  & \textbf{OA} & \textbf{mACC} & \textbf{mIOU} & \textbf{mAP} \\
\hline
DGCNN  & 88.91\% & 76.58\% & 65.34\% & 59.59\% \\
PointNet2  & 84.15\% & 71.33\% & 57.36\% & 42.48\% \\
PointMLP  & 90.25\% & 76.11\% & 66.07\% & 65.12\% \\
TSGCNet  & 90.35\% & 79.85\% & 71.09\% & 63.43\% \\
CurveNet  & 89.00\% & 77.08\% & 65.70\% & 57.56\% \\
Fast-TGCN  & 91.12\% & 81.39\% & 72.09\% & 65.79\% \\
THISNet  & 85.77\% & 68.50\% & 58.41\% & 59.46\% \\
Ours  & \textbf{91.15\%} & \textbf{81.96\%} & \textbf{73.23\%} & \textbf{70.33\%} \\
\end{tabular}
\end{table}

To further evaluate the proposed method on complex dental point clouds, additional inference experiments were conducted on a subset selected from the original validation set (125 samples). This subset specifically includes challenging cases such as missing teeth and malposed teeth, which are representative of complex dental variations. As summarized in Table~\ref{tab:comp_special}, \mname\ achieves optimal performance across all evaluation metrics. Notably, it exhibits substantial improvements in mean Average Precision (mAP)—the primary focus of our instance-level evaluation—demonstrating clear superiority over competing methods. These results confirm the effectiveness of our approach in handling complex clinical scenarios.

\subsection{Ablation study} 
To evaluate the contribution of the instance segmentation module, an ablation study was conducted by replacing it with a standard MLP-based semantic segmentation head. As shown in Table~\ref{tab:comp_ins}, removing this module leads to a noticeable decline across all evaluation metrics. This result confirms the critical role of the instance segmentation module in improving instance-aware segmentation performance. By generating discriminative embeddings for individual tooth instance masks, the module enhances the model's ability to represent and distinguish between different tooth instances. Without it, the network fails to capture sufficiently discriminative instance-level features, resulting in reduced segmentation accuracy and instance detection performance. These findings validate the effectiveness of the proposed instance segmentation module.

\begin{table}[h]
\centering
\caption{Ablation study on instance segmentation module (Ins.).}
\label{tab:comp_ins}
\begin{tabular}{c|cccc}
\textbf{Methods}  & \textbf{OA} & \textbf{mACC} & \textbf{mIOU} & \textbf{mAP} \\
\hline
W/o Ins. & 80.81\% & 59.54\% & 55.06\% & 54.14\% \\
With Ins.  & \textbf{95.20\%} & \textbf{90.19\%} & \textbf{84.42\%} & \textbf{81.93\%} \\
\end{tabular}
\end{table}

To evaluate the impact of boundary loss and graph cut on point cloud instance segmentation performance, we conducts ablation studies on these components. As summarized in Table~\ref{tab:comp_gc_bll}, the complete model achieves the best results across all metrics including OA, mAcc, mIoU, and mAP. With the boundary loss removed, mAcc shows a noticeable decline. When graph cut optimization is excluded, reductions are observed in OA, mIoU, and mAP. The configuration excluding both boundary loss and graph cut yields the lowest scores across all metrics. The ablation study results confirm the effectiveness of these two modules for achieving accurate and complete tooth instance segmentation.

\begin{table}[h]
\centering
\caption{Ablation study on boundary loss (BL) and graph cut (GC).}
\label{tab:comp_gc_bll}
\begin{tabular}{cc|cccc}
\textbf{GC} & \textbf{BL} & \textbf{OA} & \textbf{mACC} & \textbf{mIOU} & \textbf{mAP} \\
\hline
$\times$ & $\times$ & 94.09\% & 85.23\% & 79.69\% & 78.64\% \\
$\checkmark$ & $\times$  & 94.91\% & 87.40\% & 82.32\% & 79.39\% \\
$\times$ & $\checkmark$  & 94.35\% & \textbf{90.22\%} & 8 3.82\% & 79.91\% \\
$\checkmark$ & $\checkmark$  & \textbf{95.20\%} & 90.19\% & \textbf{84.42\%} & \textbf{81.93\%} \\
\end{tabular}
\end{table}

\subsection{Discussion} 
Experimental results indicate that our proposed \mname\ achieves state-of-the-art performance across all evaluation metrics, especially on challenging cases like missing and wisdom teeth. The method significantly improves both the overall accuracy and the instance-level segmentation integrity of tooth point clouds. This advantage originates from our formulation of tooth segmentation as an instance segmentation task, rather than a semantic segmentation one. The introduced instance segmentation module learns discriminative feature representations for individual tooth instances from intraoral scan data, enabling accurate generation of corresponding instance masks. Moreover, the proposed boundary-aware loss function explicitly enhances the discrimination of inter-tooth boundaries, thereby substantially improving segmentation precision in densely packed dental structures.

Future work should aim to reduce the computational complexity of the feature extraction backbone, thereby enabling the direct processing of high-resolution dental point clouds. This would eliminate the need for segmentation on simplified data, as simplification operations often result in a loss of critical detail.

\section{Conclusion} 
\label{sec:conclusion}
This paper presented \mname, a novel network architecture that formulated tooth segmentation as an instance segmentation task rather than a semantic segmentation problem. The framework designs a U-Net-shaped backbone based on PointMLP to effectively extract both local geometric details and global semantic features from intraoral scan data. A dedicated instance segmentation module is incorporated to enhance instance discrimination capability, significantly improving both overall segmentation accuracy and instance-level completeness. Additionally, a boundary-aware loss function is introduced to explicitly refine the segmentation boundaries between adjacent teeth. Comprehensive experiments validate that our method achieves state-of-the-art performance, demonstrating particular effectiveness in challenging clinical cases.
{
    \small
    \bibliographystyle{ieeenat_fullname}
    \bibliography{main}

@String(CVPR= {IEEE Conf. Comput. Vis. Pattern Recog.})

@String(TOG= {ACM Trans. Graph.})

@String(CVPR  = {CVPR})

@String(TOG   = {ACM TOG})

@inproceedings{sun2020automatic,
  title={Automatic tooth segmentation and dense correspondence of 3D dental model},
  author={Sun, Diya and Pei, Yuru and Li, Peixin and Song, Guangying and Guo, Yuke and Zha, Hongbin and Xu, Tianmin},
  booktitle={International Conference on Medical Image Computing and Computer-Assisted Intervention},
  pages={703--712},
  year={2020},
  organization={Springer}
}

@inproceedings{zhang2021tsgcnet,
  title={TSGCNet: Discriminative geometric feature learning with two-stream graph convolutional network for 3D dental model segmentation},
  author={Zhang, Lingming and Zhao, Yue and Meng, Deyu and Cui, Zhiming and Gao, Chenqiang and Gao, Xinbo and Lian, Chunfeng and Shen, Dinggang},
  booktitle={Proceedings of the IEEE/CVF Conference on Computer Vision and Pattern Recognition},
  pages={6699--6708},
  year={2021}
}

@inproceedings{sun2020tooth,
  title={Tooth segmentation and labeling from digital dental casts},
  author={Sun, Diya and Pei, Yuru and Song, Guangying and Guo, Yuke and Ma, Gengyu and Xu, Tianmin and Zha, Hongbin},
  booktitle={2020 IEEE 17th International Symposium on Biomedical Imaging (ISBI)},
  pages={669--673},
  year={2020},
  organization={IEEE}
}

@article{zhang2020automatic,
  title={Automatic 3D tooth segmentation using convolutional neural networks in harmonic parameter space},
  author={Zhang, Jianda and Li, Chunpeng and Song, Qiang and Gao, Lin and Lai, Yu-Kun},
  journal={Graphical Models},
  volume={109},
  pages={101071},
  year={2020},
  publisher={Elsevier}
}

@article{lian2020deep,
  title={Deep multi-scale mesh feature learning for automated labeling of raw dental surfaces from 3D intraoral scanners},
  author={Lian, Chunfeng and Wang, Li and Wu, Tai-Hsien and Wang, Fan and Yap, Pew-Thian and Ko, Ching-Chang and Shen, Dinggang},
  journal={IEEE transactions on medical imaging},
  volume={39},
  number={7},
  pages={2440--2450},
  year={2020},
  publisher={IEEE}
}

@inproceedings{xiong2023tsegformer,
  title={Tsegformer: 3d tooth segmentation in intraoral scans with geometry guided transformer},
  author={Xiong, Huimin and Li, Kunle and Tan, Kaiyuan and Feng, Yang and Zhou, Joey Tianyi and Hao, Jin and Ying, Haochao and Wu, Jian and Liu, Zuozhu},
  booktitle={International Conference on  Medical Image Computing and Computer-Assisted Intervention},
  pages={421--432},
  year={2023},
  organization={Springer}
}

@inproceedings{sinthanayothin2008orthodontics,
  title={Orthodontics treatment simulation by teeth segmentation and setup},
  author={Sinthanayothin, Chanjira and Tharanont, Wichit},
  booktitle={2008 5th International Conference on Electrical Engineering/Electronics, Computer, Telecommunications and Information Technology},
  volume={1},
  pages={81--84},
  year={2008},
  organization={IEEE}
}

@inproceedings{yaqi2010computer,
  title={Computer aided orthodontics treatment by virtual segmentation and adjustment},
  author={Yaqi, Ma and Zhongke, Li},
  booktitle={2010 International Conference on Image Analysis and Signal Processing},
  pages={336--339},
  year={2010},
  organization={IEEE}
}

@article{zou2015interactive,
  title={Interactive tooth partition of dental mesh base on tooth-target harmonic field},
  author={Zou, Bei-ji and Liu, Shi-jian and Liao, Sheng-hui and Ding, Xi and Liang, Ye},
  journal={Computers in Biology and Medicine},
  volume={56},
  pages={132--144},
  year={2015},
  publisher={Elsevier}
}

@inproceedings{lian2019meshsnet,
  title={Meshsnet: Deep multi-scale mesh feature learning for end-to-end tooth labeling on 3d dental surfaces},
  author={Lian, Chunfeng and Wang, Li and Wu, Tai-Hsien and Liu, Mingxia and Dur{\'a}n, Francisca and Ko, Ching-Chang and Shen, Dinggang},
  booktitle={International Conference on Medical Image Computing and Computer-Assisted Intervention},
  pages={837--845},
  year={2019},
  organization={Springer}
}

@article{hao2022toward,
  title={Toward clinically applicable 3-dimensional tooth segmentation via deep learning},
  author={Hao, J and Liao, W and Zhang, YL and Peng, J and Zhao, Z and Chen, Z and Zhou, BW and Feng, Y and Fang, B and Liu, ZZ and others},
  journal={Journal of Dental Research},
  volume={101},
  number={3},
  pages={304--311},
  year={2022},
  publisher={SAGE Publications Sage CA: Los Angeles, CA}
}

@article{zheng2022teethgnn,
  title={TeethGNN: semantic 3D teeth segmentation with graph neural networks},
  author={Zheng, Youyi and Chen, Beijia and Shen, Yuefan and Shen, Kaidi},
  journal={IEEE Transactions on Visualization and Computer Graphics},
  volume={29},
  number={7},
  pages={3158--3168},
  year={2022},
  publisher={IEEE}
}

@article{zanjani2021mask,
  title={Mask-MCNet: tooth instance segmentation in 3D point clouds of intra-oral scans},
  author={Zanjani, Farhad Ghazvinian and Pourtaherian, Arash and Zinger, Svitlana and Moin, David Anssari and Claessen, Frank and Cherici, Teo and Parinussa, Sarah and de With, Peter HN},
  journal={Neurocomputing},
  volume={453},
  pages={286--298},
  year={2021},
  publisher={Elsevier}
}

@article{cui2021tsegnet,
  title={TSegNet: An efficient and accurate tooth segmentation network on 3D dental model},
  author={Cui, Zhiming and Li, Changjian and Chen, Nenglun and Wei, Guodong and Chen, Runnan and Zhou, Yuanfeng and Shen, Dinggang and Wang, Wenping},
  journal={Medical Image Analysis},
  volume={69},
  pages={101949},
  year={2021},
  publisher={Elsevier}
}

@article{tian20223d,
  title={3D tooth instance segmentation learning objectness and affinity in point cloud},
  author={Tian, Yan and Zhang, Yujie and Chen, Wei-Gang and Liu, Dongsheng and Wang, Huiyan and Xu, Huayi and Han, Jianfeng and Ge, Yiwen},
  journal={ACM Transactions on Multimedia Computing, Communications, and Applications (TOMM)},
  volume={18},
  number={4},
  pages={1--16},
  year={2022},
  publisher={ACM New York, NY}
}

@inproceedings{qiu2022darch,
  title={Darch: Dental arch prior-assisted 3d tooth instance segmentation with weak annotations},
  author={Qiu, Liangdong and Ye, Chongjie and Chen, Pei and Liu, Yunbi and Han, Xiaoguang and Cui, Shuguang},
  booktitle={Proceedings of the IEEE/CVF Conference on Computer Vision and Pattern Recognition},
  pages={20752--20761},
  year={2022}
}

@inproceedings{hu2020jsenet,
  title={Jsenet: Joint semantic segmentation and edge detection network for 3d point clouds},
  author={Hu, Zeyu and Zhen, Mingmin and Bai, Xuyang and Fu, Hongbo and Tai, Chiew-lan},
  booktitle={European Conference on Computer Vision},
  pages={222--239},
  year={2020},
  organization={Springer}
}

@inproceedings{marin2019efficient,
  title={Efficient segmentation: Learning downsampling near semantic boundaries},
  author={Marin, Dmitrii and He, Zijian and Vajda, Peter and Chatterjee, Priyam and Tsai, Sam and Yang, Fei and Boykov, Yuri},
  booktitle={Proceedings of the IEEE/CVF international conference on computer vision},
  pages={2131--2141},
  year={2019}
}

@article{xu20183d,
  title={3D tooth segmentation and labeling using deep convolutional neural networks},
  author={Xu, Xiaojie and Liu, Chang and Zheng, Youyi},
  journal={IEEE Transactions on Visualization and Computer Graphics},
  volume={25},
  number={7},
  pages={2336--2348},
  year={2018},
  publisher={IEEE}
}

@article{yuan2020tooth,
  title={Tooth segmentation and gingival tissue deformation framework for 3D orthodontic treatment planning and evaluating},
  author={Yuan, Tianran and Wang, Yimin and Hou, Zhiwei and Wang, Jun},
  journal={Medical \& Biological Engineering \& Computing},
  volume={58},
  number={10},
  pages={2271--2290},
  year={2020},
  publisher={Springer}
}

@inproceedings{jain2023oneformer,
  title={Oneformer: One transformer to rule universal image segmentation},
  author={Jain, Jitesh and Li, Jiachen and Chiu, Mang Tik and Hassani, Ali and Orlov, Nikita and Shi, Humphrey},
  booktitle={Proceedings of the IEEE/CVF Conference on Computer Vision and Pattern Recognition},
  pages={2989--2998},
  year={2023}
}

@inproceedings{cheng2022masked,
  title={Masked-attention mask transformer for universal image segmentation},
  author={Cheng, Bowen and Misra, Ishan and Schwing, Alexander G and Kirillov, Alexander and Girdhar, Rohit},
  booktitle={Proceedings of the IEEE/CVF Conference on Computer Vision and Pattern Recognition},
  pages={1290--1299},
  year={2022}
}

@article{vaswani2017attention,
  title={Attention is all you need},
  author={Vaswani, Ashish and Shazeer, Noam and Parmar, Niki and Uszkoreit, Jakob and Jones, Llion and Gomez, Aidan N and Kaiser, {\L}ukasz and Polosukhin, Illia},
  journal={Advances in Neural Information Processing Systems},
  volume={30},
  year={2017}
}

@article{ben20233dteethseg,
  title={3DTeethSeg'22: 3D Teeth Scan Segmentation and Labeling Challenge},
  author={Ben-Hamadou, Achraf and Smaoui, Oussama and Rekik, Ahmed and Pujades, Sergi and Boyer, Edmond and Lim, Hoyeon and Kim, Minchang and Lee, Minkyung and Chung, Minyoung and Shin, Yeong-Gil and others},
  journal={arXiv preprint arXiv:2305.18277},
  year={2023}
}

@article{li2023thisnet,
  title={THISNet: Tooth Instance Segmentation on 3D Dental Models via Highlighting Tooth Regions},
  author={Li, Pengcheng and Gao, Chenqiang and Liu, Fangcen and Meng, Deyu and Yan, Yan},
  journal={IEEE Transactions on Circuits and Systems for Video Technology},
  volume={34},
  number={7},
  pages={5229--5241},
  year={2023},
  publisher={IEEE}
}

@article{ma2022rethinking,
  title={Rethinking Network Design and Local Geometry in Point Cloud: A Simple Residual MLP Framework},
  author={Ma, Xu and Qin, Can and You, Haoxuan and Ran, Haoxi and Fu, Yun},
  journal={arXiv preprint arXiv:2202.07123},
  year={2022}
}

@article{qi2017pointnet++,
  title={Pointnet++: Deep hierarchical feature learning on point sets in a metric space},
  author={Qi, Charles Ruizhongtai and Yi, Li and Su, Hao and Guibas, Leonidas J},
  journal={Advances in Neural Information Processing Systems},
  volume={30},
  year={2017}
}

@article{muzahid2020curvenet,
  title={CurveNet: Curvature-based multitask learning deep networks for 3D object recognition},
  author={Muzahid, AAM and Wan, Wanggen and Sohel, Ferdous and Wu, Lianyao and Hou, Li},
  journal={IEEE/CAA Journal of Automatica Sinica},
  volume={8},
  number={6},
  pages={1177--1187},
  year={2020},
  publisher={IEEE}
}

@article{li2024fine,
  title={A fine-grained orthodontics segmentation model for 3D intraoral scan data},
  author={Li, Juncheng and Cheng, Bodong and Niu, Najun and Gao, Guangwei and Ying, Shihui and Shi, Jun and Zeng, Tieyong},
  journal={Computers in Biology and Medicine},
  volume={168},
  pages={107821},
  year={2024},
  publisher={Elsevier}
}

@article{wang2019dynamic,
  title={Dynamic graph cnn for learning on point clouds},
  author={Wang, Yue and Sun, Yongbin and Liu, Ziwei and Sarma, Sanjay E and Bronstein, Michael M and Solomon, Justin M},
  journal={ACM Transactions on Graphics (tog)},
  volume={38},
  number={5},
  pages={1--12},
  year={2019},
  publisher={Acm New York, NY, USA}
}

@article{chen2017deeplab,
  title={Deeplab: Semantic image segmentation with deep convolutional nets, atrous convolution, and fully connected crfs},
  author={Chen, Liang-Chieh and Papandreou, George and Kokkinos, Iasonas and Murphy, Kevin and Yuille, Alan L},
  journal={IEEE Transactions on Pattern Analysis and Machine Intelligence},
  volume={40},
  number={4},
  pages={834--848},
  year={2017},
  publisher={IEEE}
}

@inproceedings{cai2017pancreas,
  title={Pancreas segmentation in MRI using graph-based decision fusion on convolutional neural networks},
  author={Cai, Jinzheng and Lu, Le and Xie, Yuanpu and Xing, Fuyong and Yang, Lin},
  booktitle={International Conference on Medical Image Computing and Computer-Assisted Intervention},
  pages={674--682},
  year={2017},
  organization={Springer}
}

@article{liu2022hierarchical,
  title={Hierarchical self-supervised learning for 3D tooth segmentation in intra-oral mesh scans},
  author={Liu, Zuozhu and He, Xiaoxuan and Wang, Hualiang and Xiong, Huimin and Zhang, Yan and Wang, Gaoang and Hao, Jin and Feng, Yang and Zhu, Fudong and Hu, Haoji},
  journal={IEEE Transactions on Medical Imaging},
  volume={42},
  number={2},
  pages={467--480},
  year={2022},
  publisher={IEEE}
}

@article{kuhn2005hungarian,
  author = {H. W. Kuhn},
  title = {The Hungarian method for the assignment problem},
  journal = {Nav. Res. Logistics (NRL)},
  volume = {52},
  number = {1},
  pages = {7--21},
  year = {2005},
  month = {Feb.}
}

@inproceedings{sudre2017generalised,
  author = {C. H. Sudre and W. Li and T. Vercauteren and S. Ourselin and M. J. Cardoso},
  title = {Generalised Dice overlap as a deep learning loss function for highly unbalanced segmentations},
  booktitle = {Proc. Int. MICCAI Workshop Deep Learn. Med. Image Anal. Multimodal Learn. Clin. Decis. Support (Lecture Notes in Computer Science)},
  volume = {10553},
  pages = {240--248},
  year = {2017},
  publisher = {Springer}
}

@inproceedings{jana2023critical,
  author = {Jana, A. and Maiti, A. and Metaxas, D. N.},
  title = {A Critical Analysis of the Limitation of Deep Learning based 3D Dental Mesh Segmentation Methods in Segmenting Partial Scans},
  booktitle = {Annu Int Conf IEEE Eng Med Biol Soc},
  year = {2023},
  month = {Jul},
  volume = {2023},
  pages = {1--7}
}

@inproceedings{wang2024weakly,
  author = {Wang, H. and Li, K. and Zhu, J. and Wang, F. and Lian, C. and Ma, J.},
  title = {Weakly Supervised Tooth Instance Segmentation on 3D Dental Models with Multi-label Learning},
  booktitle = {Medical Image Computing and Computer Assisted Intervention – MICCAI 2024},
  series = {Lecture Notes in Computer Science},
  volume = {15009},
  year = {2024},
  publisher = {Springer, Cham},
  editor = {Linguraru, M. G. and others}
}

@article{Xi20253DDM,
  title={3D Dental Model Segmentation with Geometrical Boundary Preserving},
  author={Shufan Xi and Zexian Liu and Junlin Chang and Hongyu Wu and Xiaogang Wang and Aimin Hao},
  journal={2025 IEEE/CVF Conference on Computer Vision and Pattern Recognition (CVPR)},
  year={2025},
  pages={10476-10485},
  url={https://api.semanticscholar.org/CorpusID:277452483}
}

@INPROCEEDINGS{10654940,
  author={Zou, Bo and Wang, Shaofeng and Liu, Hao and Sun, Gaoyue and Wang, Yajie and Zuo, FeiFei and Quan, Chengbin and Zhao, Youjian},
  booktitle={2024 IEEE/CVF Conference on Computer Vision and Pattern Recognition (CVPR)}, 
  title={Teeth-SEG: An Efficient Instance Segmentation Framework for Orthodontic Treatment Based on Multi-Scale Aggregation and Anthropic Prior Knowledge}, 
  year={2024},
  volume={},
  number={},
  pages={11601-11610},
  doi={10.1109/CVPR52733.2024.01102}}

@INPROCEEDINGS{10230650,
  author={Jana, Ananya and Subhash, Hrebesh Molly and Metaxas, Dimitris},
  booktitle={2023 IEEE 20th International Symposium on Biomedical Imaging (ISBI)}, 
  title={3D Tooth Mesh Segmentation with Simplified Mesh Cell Representation}, 
  year={2023},
  volume={},
  number={},
  pages={1-5},
  doi={10.1109/ISBI53787.2023.10230650}}
}


\end{document}